\documentclass[fleqn,usenatbib]{mnras}

\usepackage{newtxtext,newtxmath}

\usepackage[T1]{fontenc}
\usepackage{ae,aecompl}

\usepackage{color}

\usepackage{graphicx}	
\usepackage{amsmath}	
\usepackage{amssymb}	
\usepackage{upgreek}

\title[LOFAR follow-up of GW170817]{LOFAR 144-MHz follow-up observations of GW170817}

\author[J. W. Broderick et al.]{J. W. Broderick,$^{1,2}$\thanks{E-mail: jess.broderick@curtin.edu.au}
T. W. Shimwell,$^{1,3}$
K. Gourdji,$^{4}$
A. Rowlinson,$^{1,4}$ \newauthor
S. Nissanke,$^{5,6}$ 
K. Hotokezaka,$^{7,8}$
P. G. Jonker,$^{9,10}$
C. Tasse,$^{11,12}$
M. J. Hardcastle,$^{13}$ \newauthor
J. B. R. Oonk,$^{14,1,3}$
R. P. Fender,$^{15}$ 
R. A. M. J. Wijers,$^{4}$
A. Shulevski,$^{4}$\newauthor
A. J. Stewart,$^{16}$
S. ter Veen,$^{1}$
V. A. Moss,$^{17,1,16}$
M. H. D. van der Wiel,$^{1}$\newauthor
D. A. Nichols,$^{5,18}$
A. Piette,$^{19}$
M. E. Bell,$^{20}$
D. Carbone,$^{21}$
S. Corbel,$^{22,23}$\newauthor
J. Eisl{\"o}ffel,$^{24}$
J.-M. Grie{\ss}meier,$^{25,23}$
E. F. Keane,$^{26}$
C. J. Law,$^{27}$\newauthor
T. Mu{\~n}oz-Darias,$^{28,29}$
M. Pietka,$^{15}$
M. Serylak,$^{30,31}$
A. J. van der Horst,$^{32,33}$\newauthor
J. van Leeuwen,$^{1,4}$
R. Wijnands,$^{4}$
P. Zarka,$^{34}$
J. M. Anderson,$^{35,36}$\newauthor 
M. J. Bentum,$^{1,37}$
R. Blaauw,$^{1}$
W. N. Brouw,$^{38}$
M. Br{\"u}ggen,$^{39}$
B. Ciardi,$^{40}$\newauthor 
M. de Vos,$^{1}$ 
S. Duscha,$^{1}$
R. A. Fallows,$^{1}$
T. M. O. Franzen,$^{1}$
M. A. Garrett,$^{41,3}$\newauthor 
A. W. Gunst,$^{1}$
M. Hoeft,$^{24}$
J. R. H{\"o}randel,$^{10,6,42}$
M. Iacobelli,$^{1}$
E. J{\"u}tte,$^{43}$\newauthor 
L. V. E. Koopmans,$^{38}$
A. Krankowski,$^{44}$
P. Maat,$^{1}$
G. Mann,$^{45}$
H. Mulder,$^{1}$\newauthor 
A. Nelles,$^{46,47}$
H. Paas,$^{48}$
M. Pandey-Pommier,$^{23,49}$ 
R. Pekal,$^{50}$
W. Reich,$^{51}$\newauthor 
H. J. A. R{\"o}ttgering,$^{3}$
D. J. Schwarz,$^{52}$
O. Smirnov,$^{12,30}$
M. Soida,$^{53}$\newauthor 
M. C. Toribio,$^{54}$
M. P. van Haarlem,$^{1}$ 
R. J. van Weeren,$^{3}$
C. Vocks,$^{45}$\newauthor 
O. Wucknitz$^{51}$ and
P. Zucca$^{1}$ 
\\
\\
\large{Affiliations are listed at the end of the paper}
}
\date{Accepted XXX. Received YYY; in original form ZZZ}

\pubyear{2020}

\begin{document}
\label{firstpage}
\pagerange{\pageref{firstpage}--\pageref{lastpage}}
\maketitle

\begin{abstract}
We present low-radio-frequency follow-up observations of AT~2017gfo, the electromagnetic counterpart of GW170817, which was the first binary neutron star merger to be detected by Advanced LIGO--Virgo. These data, with a central frequency of 144 MHz, were obtained with LOFAR, the Low-Frequency Array. The maximum elevation of the target is just 13\fdg7 when observed with LOFAR, making our observations particularly challenging to calibrate and significantly limiting the achievable sensitivity. On time-scales of 130--138 and 371--374 days after the merger event, we obtain 3$\sigma$ upper limits for the afterglow component of 6.6 and 19.5 mJy beam$^{-1}$, respectively. Using our best upper limit and previously published, contemporaneous higher-frequency radio data, we place a limit on any potential steepening of the radio spectrum between 610 and 144 MHz: the two-point spectral index $\alpha^{610}_{144} \gtrsim$~$-$2.5. We also show that LOFAR can detect the afterglows of future binary neutron star merger events occurring at more favourable elevations.
\end{abstract}

\begin{keywords}
gravitational waves -- radio continuum: stars -- stars: neutron
\end{keywords}



\section{Introduction}\label{section:introduction}

On 2017 August 17, a network comprising the Advanced Laser Interferometer Gravitational-Wave Observatory and the Advanced Virgo interferometer \citep[Advanced LIGO--Virgo;][]{acernese15,aasi15} detected gravitational waves (GWs) from the binary neutron star merger GW170817 \citep[][]{abbott17c}. The subsequent discovery and initial characterisation of the electromagnetic counterpart, AT~2017gfo (\citealt[][]{abbott17a} and references therein), located in the galaxy NGC~4993 (heliocentric redshift $z =$ 0.00978; distance $\approx$ 40 Mpc; \citealt[][]{hjorth17}), was truly a landmark event in multimessenger astrophysics.  

Following the short gamma-ray burst (sGRB) associated with this event, GRB~170817A \citep[][]{abbott17a,abbott17b,goldstein17}, radio emission was anticipated as the associated merger outflow interacted with the circum-merger medium. Monitoring the radio emission could therefore provide crucial information on the energetics and geometry of the outflow, as well as the ambient environment. At radio frequencies, telescopes were observing the Advanced LIGO--Virgo probability region for GW170817 within 29 min post merger \citep[][]{callister17a}, and subsequent monitoring of AT~2017gfo resulted in an initial radio detection 16 days after the event \citep{abbott17a,hallinan17}. Further monitoring, predominantly at frequencies between 0.6 and 15 GHz, has since taken place \citep[e.g.][]{alexander17,alexander18,mooley18a,mooley18b,mooley18c,margutti18,dobie18,troja18,troja19,corsi18,resmi18}. At these frequencies, a general picture emerged in which the radio light curve was first observed to steadily rise, before it turned over and began a more rapid decay. Using a compilation of 0.6--10 GHz radio data from 17--298 days post merger, \citet[][]{mooley18c} derived both a fitted time for the radio peak of 174$^{+9}_{-6}$ days and a fitted 3-GHz peak flux density of 98$^{+8}_{-9}$ $\upmu$Jy (also see similar analyses in \citealt[][]{dobie18} and \citealt[][]{alexander18}). The fitted radio spectral index $\alpha$\footnote{We use the convention that $S_{\nu} \propto \nu^{\alpha}$, where $S_{\nu}$ is the flux density at frequency $\nu$.} from this study is $-$0.53 $\pm$ 0.04, consistent with broadband spectral indices determined using radio, optical and X-ray data at various epochs, where the typical value is approximately $-$0.58 \citep[e.g.][]{margutti18,troja18,troja19,alexander18,hajela19}. \citet[][]{mooley18c} also found power-law dependencies for the rise and decay phases of approximately $t^{0.8}$ and $t^{-2.4}$, respectively, where $t$ is the time since the merger. Within the associated uncertainties, these results are consistent with the broadband evolution of AT~2017gfo \citep[e.g.][]{alexander18,lamb19,troja19,hajela19}. 

Two competing models emerged to explain the radio light curve: either the jet successfully broke through the surrounding cocoon of ejected material (also known as a `structured' jet) but was observed off-axis, or the jet was `choked' by the cocoon, in which it deposited all of its energy \citep[e.g.][]{troja17,kasliwal17}. The observed evolution of the radio light curve, and very-long-baseline interferometric measurements of both apparent superluminal motion and a sufficiently compact apparent source size, confirmed that a jet was successfully launched for GW170817 with opening angle $<$5\degr\,and observed from a viewing angle of approximately 15--20\degr\,\citep{alexander18,mooley18b,mooley18c,ghirlanda19}.

Although predicted faint flux density levels and slow light curve rise times \citep[e.g.][]{hotokezaka16} may make late-time, low-radio-frequency ($\lesssim$ 200 MHz) detections challenging, such flux density measurements can help to discriminate between competing models for the radio emission following a compact binary merger. In addition, the current generation of low-frequency aperture arrays have rapid electronic beam steering, as well as very large fields of view that can cover at minimum a significant fraction of the Advanced LIGO--Virgo probability region for a GW event. Therefore, there is the interesting potential to use low-frequency aperture arrays to search for prompt, coherent radio emission from a compact binary merger (e.g. \citealt{callister19}; also see \citealt[][]{obenberger14},  \citealt[][]{yancey15}, \citealt[][]{kaplan15,kaplan16}, \citealt[][]{chu16}, \citealt[][]{anderson18}, \citealt[][]{rowlinson19}, \citealt[][]{james19} and \citealt[][]{rowlinson19b}).

At low frequencies, GW170817 was followed up with both the first station of the Long Wavelength Array \citep[LWA1;][]{ellingson13} and the Murchison Widefield Array \citep[MWA;][]{tingay13}. Details of the LWA1 observations can be found in \citet[][]{callister17a,callister17b,callister17c} and \citet[][]{abbott17a}, including the aforementioned observation 29~min post merger, as well as additional observations up to approximately 13 days after the event. Similarly, details of the MWA observations, occurring 0.8--4.9 days post merger, can be found in \citet[][]{kaplan17a,kaplan17b}, \citet[][]{abbott17a} and \citet[][]{andreoni17}. At the position of NGC~4993, 26- and 45-MHz LWA1 observations approximately 8 h after the event yielded 3$\sigma$ upper limits of 200 and 100 Jy, respectively, for persistent emission \citep[][]{callister17c}. At 185 MHz, 0.8 days post merger, the 3$\sigma$ MWA upper limit was 51 mJy beam$^{-1}$, albeit with only 40 of the 128 tiles operational at the time \citep[][]{kaplan17b, andreoni17}.

In this paper, we present late-time (130--138 and 371--374 days post merger), low-radio-frequency follow-up observations of AT~2017gfo, obtained with the high-band antennas (HBA) of the Low-Frequency Array \citep[LOFAR;][]{vanhaarlem13}. In Section~\ref{section:observations}, we describe these observations, and how the data were calibrated and imaged. Our results are presented in Section~\ref{section:results}, which is followed by a discussion in Section~\ref{section:discussion} on the additional constraints that our 144-MHz upper limits place on the properties of the radio emission from GW170817, as well as future prospects for LOFAR when observing new GW events. We then conclude in Section~\ref{section:conclusions}. All uncertainties reported in this paper are quoted at the 68 per cent confidence level.        

\section{LOFAR observations and data reduction}\label{section:observations}

Table~\ref{table:observations} presents a log of the two observing runs. In both cases, we used the `HBA Dual Inner' configuration, with 24 core stations and either 13 or 14 Dutch remote stations. Because the location of AT~2017gfo on the sky (Dec $=$ $-$23\fdg4) is very far south relative to LOFAR (latitude of core $=$ 52\fdg9 N), both runs were split into 4 $\times$ 2-h observations on separate days, each centred as closely to transit as possible, so as to maximize the elevation of the target. Nonetheless, the maximum elevation as viewed from the LOFAR core is only 13\fdg7, which significantly affected the sensitivity that we could achieve due to the small projected station area, as well as making our observations far more susceptible to ionospheric effects. Both sets of observations comprised 380 $\times$ 195.3-kHz sub-bands that spanned the frequency range 115--189 MHz, although in this study we made use of the most sensitive part of the bandpass between 120--168 MHz (246 sub-bands). All 2-h observations were preceded by a 10-min scan of the flux density calibrator 3C~295.        

Preprocessing consisted of flagging and averaging (in time and/or frequency) steps. The former made use of {\sc aoflagger} \citep*[][]{offringa10,offringa12a,offringa12b}, with 14.3 and 15.9 per cent of data flagged per sub-band on average for Runs 1 and 2, respectively. After preprocessing, the temporal and frequency resolutions for Run 1 were 1 s and 12.2 kHz (i.e. 16 channels per sub-band), respectively; the corresponding values were 4 s and 48.8 kHz (i.e. 4 channels per sub-band), respectively, for Run 2. These differing resolutions between Runs 1 and 2 were due to the fact that Run 2 was obtained as part of a larger LOFAR GW follow-up project with different preprocessing settings (Gourdji et al. in prep.). In principle, however, Run 2 still had sufficiently fine temporal and frequency resolutions to permit proper calibration, as was the case for Run 1 \citep[e.g. see][]{shimwell17,shimwell19}.     

The reduction steps after preprocessing were direction-independent calibration, followed by direction-dependent calibration and imaging. A detailed description of the procedure, which was developed for the LOFAR Two-metre Sky Survey (LoTSS), can be found in \citet[][]{shimwell17,shimwell19}. The direction-independent pipeline removed the effects outlined in \citet[][]{degasperin19}, following the procedure described in \citet[][]{vanweeren16} and \citet[][]{williams16}; it made use of the {\sc bbs} \citep[][]{pandey09} and {\sc dppp} \citep[][]{vandiepen18} software packages.\footnote{\url{https://github.com/lofar-astron/prefactor} using commit dd68c57.} We used a model for 3C~295 that is consistent with the \citet[][]{scaife12} flux density scale. The sky model used to calibrate the target data was derived from the TIFR Giant Metrewave Radio Telescope \citep[GMRT;][]{swarup91} Sky Survey First Alternative Data Release \citep[TGSS ADR1;][]{intema17}. 

The direction-dependent step made use of {\sc kms} \citep[][]{tasse14a,tasse14b,smirnov15} and {\sc ddfacet} \citep[][]{tasse18} for calibration and imaging, respectively.\footnote{In particular, we used version 2.2 of {\sc ddf-pipeline}, which can be found at \url{https://github.com/mhardcastle/ddf-pipeline}.}  The calibrated data per 2-h observation comprised 25 blocks of 10 sub-bands each; the highest-frequency block had four empty sub-bands because we used 246 sub-bands in total. The final temporal and frequency resolutions were 8 s and 97.7 kHz (i.e. 2 channels per sub-band), respectively, and the central frequency was 144 MHz.          

\begin{table}
 \centering
  \caption{An observing log for our LOFAR observations of AT~2017gfo. Both observations had a frequency range of 120--168 MHz, and a central frequency of 144 MHz.}
  \begin{tabular}{ccc}
  \hline
 &  \multicolumn{2}{c}{Observing run}  \\ 
 & 1 & 2  \\
 \hline
 Date & 2017 December 25, 27,  & 2018 August  \\
 & 28, 2018 January 2  & 23--26 \\
\\
 Days post merger & 130--138 & 371--374 \\
\\
 RMS noise level & 2.1 $\pm$ 0.6 & 6.2 $\pm$ 1.9 \\
(mJy beam$^{-1}$) & & \\
\\
Observation IDs & 632609-- & 664578-- \\
& 632637$^{\rm a}$ & 664604$^{\rm b}$ \\
\hline
\multicolumn{3}{l}{$^{\rm a}$ IDs increase in steps of 4.}\\
\multicolumn{3}{l}{$^{\rm b}$ IDs increase in alternate steps of 2 and 6.}\\
\end{tabular}
\label{table:observations}
\end{table}

\section{Results}\label{section:results}

\begin{figure*}
\begin{minipage}{0.75\textwidth}
\includegraphics[height=10cm]{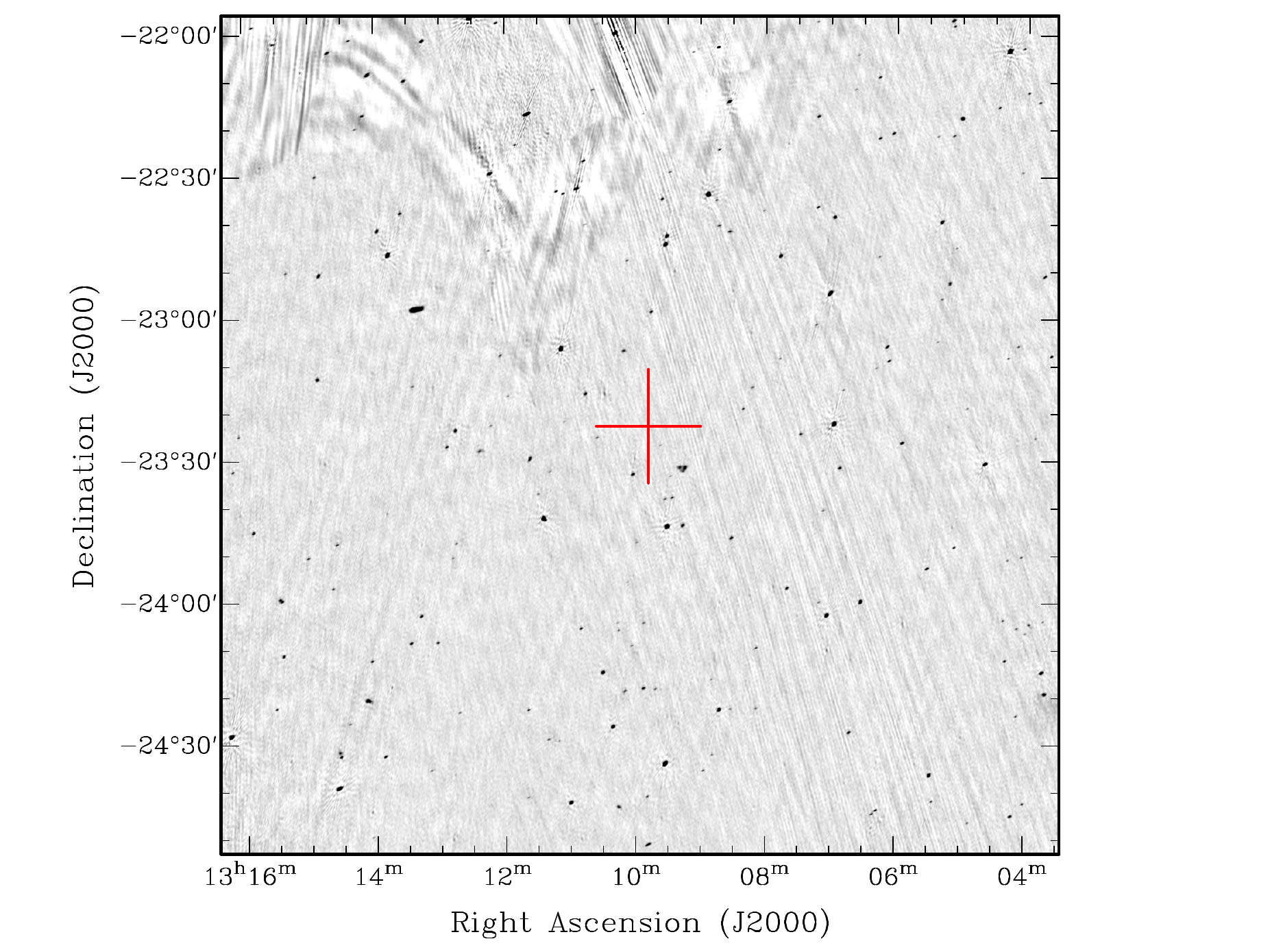}
\end{minipage}%
\begin{minipage}{0.25\textwidth}
\vspace*{-0.7cm}
\hspace*{-2.cm}\includegraphics[height=4.7cm]{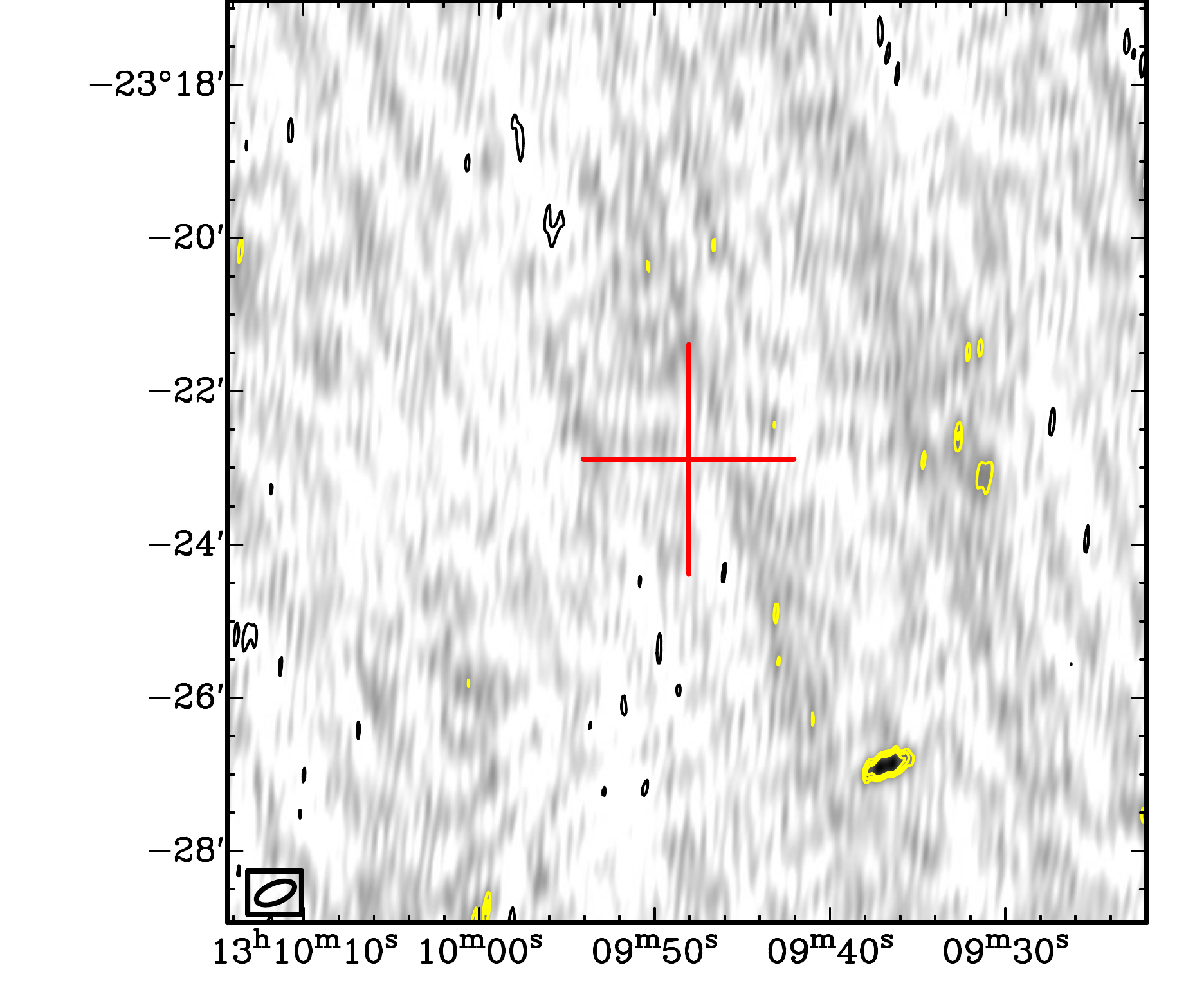}
\vspace*{-0.2cm}
\hspace*{-2.cm}\includegraphics[height=4.7cm]{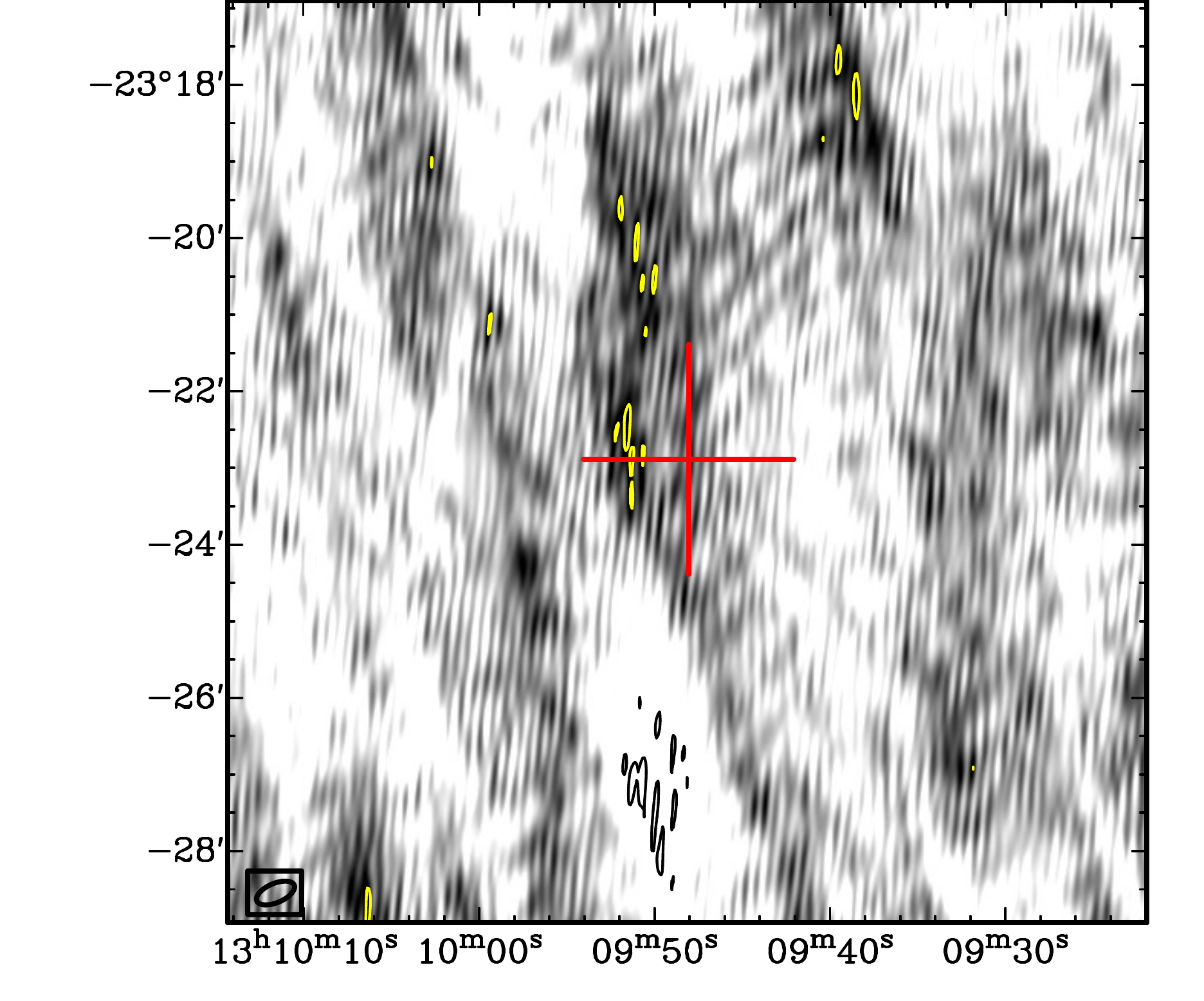}
\end{minipage}
\vspace{-0.1cm}
\caption{The left panel is our LOFAR 144-MHz image of a field centred on AT~2017gfo, obtained from the first set of observations 130--138 days post merger. We show a subsection with dimensions 3\degr\,$\times$ 3\degr; the crosshairs mark the position of AT~2017gfo. The two panels on the right (with the same grey-scale contrast scheme) show a more zoomed-in view (0.2\degr \,$\times$ 0.2\degr), where Run 1 is the top panel and Run 2 the bottom panel. The RMS noise levels in the vicinity of the position of AT~2017gfo are approximately 2.1 and 6.2 mJy beam$^{-1}$ for Runs 1 and 2, respectively; the contour scheme is $-$3$\sigma$ (black) and (3, 4, 5) $\times$ $\sigma$ (yellow). The restoring beam, chosen to be the same for both runs (32 arcsec $\times$ 15 arcsec; beam position angle $-$67\degr\,measured north through east) is shown in the bottom left-hand corner of these two panels.}
\label{fig:lofar_map}
\end{figure*}

In the left panel of Figure~\ref{fig:lofar_map}, we show our LOFAR map from Run 1, centred on the position of AT~2017gfo. The corresponding map from Run 2 has a lower dynamic range. When imaging, we had to take into account a primary beam that is very elongated in the north-south direction (FWHM $\approx$ 23\fdg6 $\times$ 4\fdg8). The angular resolution is as good as 15 arcsec, with full details of the synthesised beams given in Figure~\ref{fig:lofar_map}. 

At the angular resolution of our observations, any emission from AT~2017gfo and the active galactic nucleus of NGC~4993 could be partially blended, depending on the brightnesses of the sources. In neither of our maps is this potentially blended emission detected, nor are there detections at the two separate sets of coordinates (right panels of Figure~\ref{fig:lofar_map}). In terms of establishing upper limits for the target flux density, the largest source of uncertainty is related to a standard frequency-dependent error in the flux density scale (see references on LOFAR calibration in Section~\ref{section:observations} for further details), which we have corrected for, to first order, by bootstrapping to TGSS within a 1\degr\,radius from the position of AT~2017gfo. We used an integrated flux density to peak flux density ratio of $\leq$ 1.5 in TGSS to restrict the bootstrapping procedure to sources within the search radius that were not too extended, while retaining a sufficient number of sources for a reasonable statistical comparison. A small adjustment was also made for the slightly different central frequency of TGSS (147.5 MHz), assuming a canonical spectral index of $-$0.7. Source finding in the LOFAR maps made use of {\sc pybdsf} \citep[][]{mohan15}. 

The multiplicative correction factors to apply to our LOFAR maps were found to be 1.3 $\pm$ 0.3 and 3.3 $\pm$ 1.0 for Runs 1 and 2, respectively. Moreover, in this region of sky, we found that the TGSS flux density scale is consistent to within about 10 per cent on average with the corresponding scale from the Galactic and Extragalactic All-sky Murchison Widefield Array Survey \citep[GLEAM;][]{hurleywalker17}. Using appropriate error propagation, we combined each bootstrapping uncertainty with a 10 per cent absolute flux density calibration uncertainty. After doing this, the RMS noise levels ($\sigma$) are 2.1 $\pm$ 0.6 and 6.2 $\pm$ 1.9 mJy beam$^{-1}$ for Runs 1 and 2, respectively, in the vicinity of the target position. To obtain 3$\sigma$ upper limits that are correct to first order, we then combined each RMS value with its respective uncertainty, in quadrature, before multiplying by three. Therefore, our 3$\sigma$ upper limits for AT~2017gfo (and NGC~4993) are 6.6 and 19.5 mJy beam$^{-1}$ for Runs 1 and 2, respectively. 

As is apparent in Figure~\ref{fig:lofar_map}, the dynamic range is limited in the north and north-east of the map, but the image quality nearer the centre of the map is relatively unaffected. Moreover, to first order, there is a good correspondence between the source morphologies and positions in the LOFAR and TGSS images.\footnote{The TGSS image archive can be found at \url{https://vo.astron.nl/tgssadr/q_fits/imgs/info}.} While a detailed comparison is beyond the scope of this paper, it is worth noting that the noise level reported above for Run 1 is the deepest value obtained thus far in the literature for LOFAR interferometric observations significantly south of the celestial equator. Unfortunately, however, the noisier map from the second run does not share the same overall consistency with TGSS, and the correction factor reported above for Run 2 is unusually large. Poorer ionospheric conditions are likely to be a contributing factor. The upper limit from this run should therefore be viewed with caution, although we note that this does not affect any subsequent discussion in this paper.      

\section{Discussion}\label{section:discussion}

\begin{figure*}  
\begin{minipage}{0.5\textwidth}
\includegraphics[height=6.25cm]{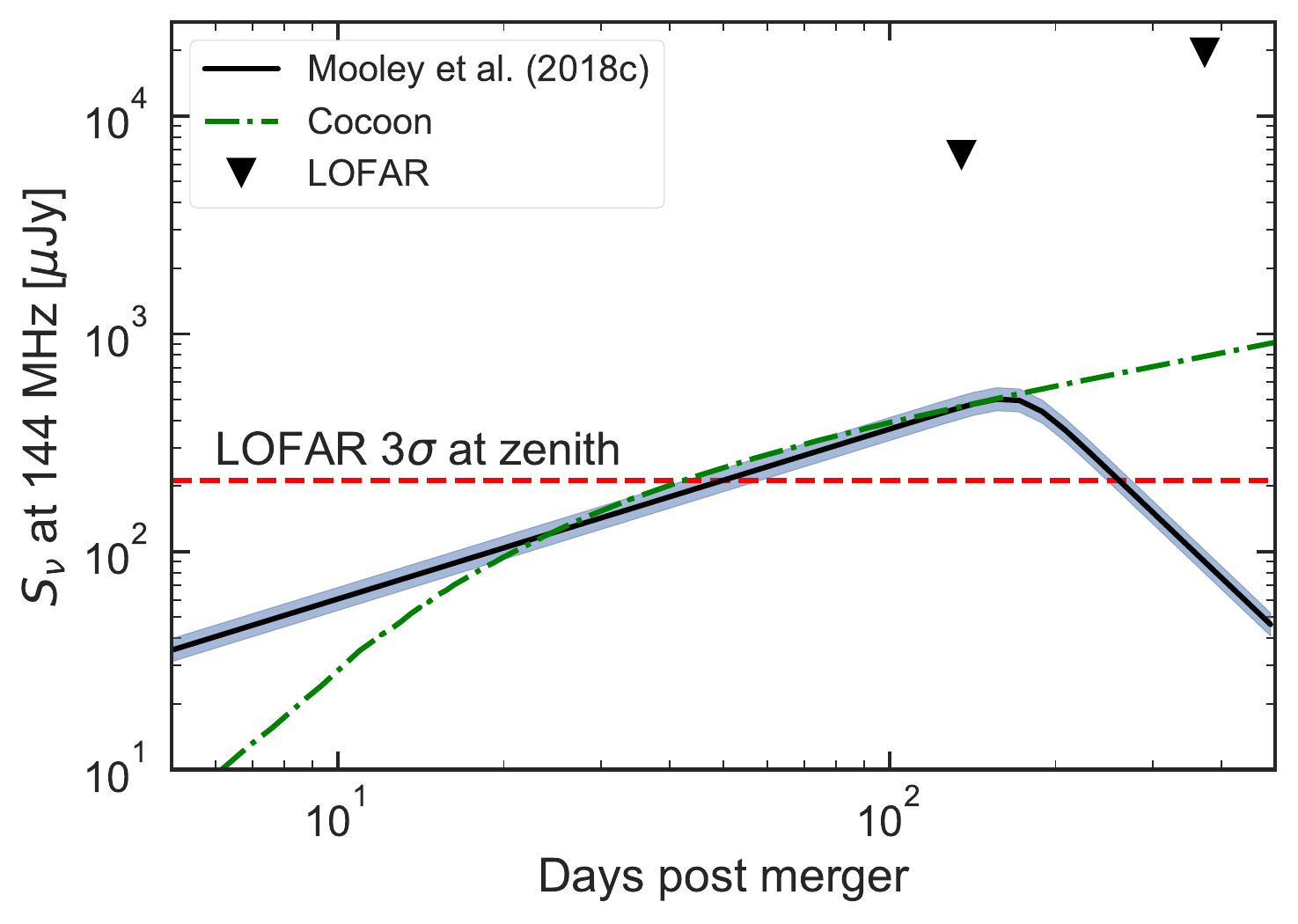}
\end{minipage}%
\begin{minipage}{0.5\textwidth}
\includegraphics[height=6.25cm]{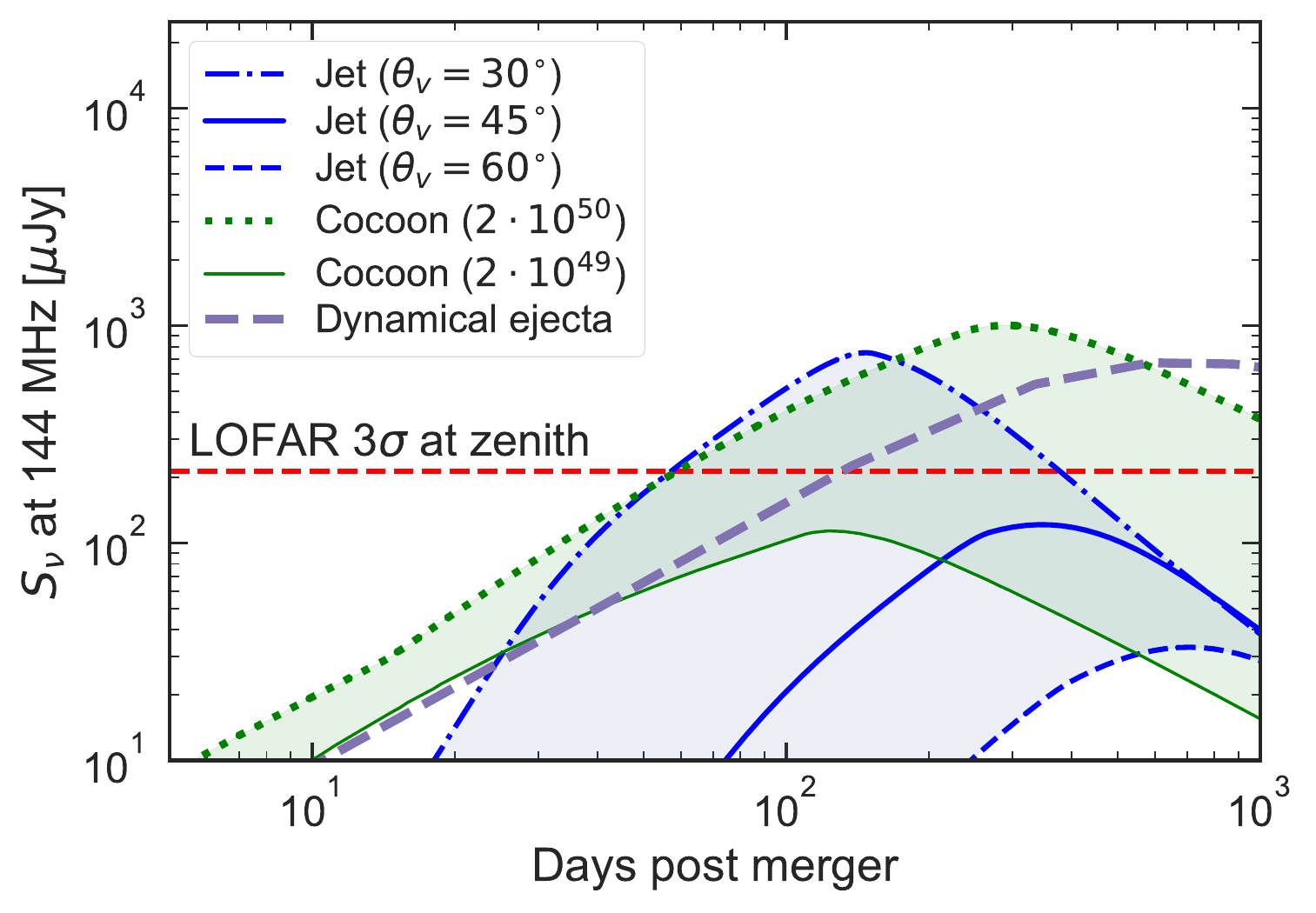}
\end{minipage}
\caption{Afterglow light curves at 144 MHz for GW170817 (left panel) and an example future event at a distance of 100 Mpc (right panel). In the left panel, the solid line with surrounding shading is the light curve at 144 MHz extrapolated from the observations at higher frequencies with $\alpha =$ $-$0.53 $\pm$ 0.04 \citep[][]{mooley18c}. The triangles are our 3$\sigma$ LOFAR upper limits, and the median 3$\sigma$ sensitivity of LOFAR for routine 8-h observations for declinations at or near zenith (see Section~\ref{section:discussion2}) is depicted as a dashed horizontal line. The dash-dotted line is an analytic cocoon model described in \citet[][]{mooley18a} with circum-merger density $n=$ 10$^{-4}$ cm$^{-3}$, and microphysical parameters $\epsilon_{e}=$ 0.1, $\epsilon_{B}=$ 0.01, and $p=$ 2.1. In the right panel, $n$ is chosen to be 0.01 cm$^{-3}$, and the microphysical parameters are $\epsilon_{e}=$ 0.1, $\epsilon_{B}=$ 0.01, and $p=$ 2.2. We show a structured jet model for various viewing angles, details of which are given in Section~\ref{section:discussion2}. For the cocoon model in this panel, we show two different kinetic energies of 2 $\times$ 10$^{49}$ and 2 $\times$ 10$^{50}$ erg. A dynamical ejecta light curve (model `DNS$_{\rm m}$') taken from \citet[][]{hotokezaka16} is also shown.
}
\label{fig:modelling}
\end{figure*}

\subsection{Low-frequency constraints on the radio spectrum of AT~2017gfo}\label{section:discussion1} 

We now discuss the additional constraints that can be placed on the radio spectrum of AT~2017gfo, as well as NGC~4993. First, \citet[][]{resmi18} presented 610- and 1390-MHz GMRT flux densities for both AT~2017gfo and the nucleus of NGC~4993. In the case of NGC~4993, the flux densities are relatively faint, and the radio spectrum relatively flat. After averaging the reported flux densities using inverse-variance weighting, we find that $\overline{S}_{610} \approx$ 0.99 mJy, $\overline{S}_{1390} \approx$ 0.78 mJy, and the mean two-point spectral index $\overline{\alpha}^{1390}_{610} \approx$ $-$0.29. Therefore, our best 3$\sigma$ upper limit at 144 MHz only provides a very weak, additional constraint of $\alpha^{610}_{144} \gtrsim$ $-$1.3.     

In the case of AT~2017gfo, our first set of observations is either bookended by or close in time to a selection of the 610- and 1390-MHz GMRT observations, as well as Australia Telescope Compact Array \citep*[ATCA;][]{frater92,wilson11} observations carried out at 5500 and 9000 MHz \citep[][]{dobie18}. For the 1390-MHz observations that bracketed our data, we averaged the corresponding flux densities using inverse-variance weighting; we also did this for the ATCA data, but using the 7250-MHz flux densities with a correction factor applied from \citet[][]{mooley18c}. We can then combine these (averaged) flux densities with our LOFAR 3$\sigma$ upper limit from Run 1 to calculate approximate constraints on a number of two-point spectral indices, roughly 125--150 days post merger. The constraints are $\alpha^{610}_{144} \gtrsim$ $-$2.5, $\alpha^{1390}_{144} \gtrsim$ $-$1.8, and  $\alpha^{7250}_{144} \gtrsim$ $-$1.2. These limits are still significantly steeper than the fitted 0.6--10 GHz radio spectral index of $-$0.53 $\pm$ 0.04 as determined by \citet[][]{mooley18c} (see Section~\ref{section:introduction}). Therefore, we can only rule out that the radio spectrum of AT~2017gfo does not steepen below 610 MHz to an extreme degree. 

Coherent radio emission can result in an ultra-steep spectrum component that is only observable at low frequencies. An overview of the physical mechanisms by which coherent radio emission may arise from a compact binary merger was given by \citet[][]{rowlinson19}. In the case of GW170817, and on a time-scale of 130--138 days after the merger, there are two immediate considerations. First, two-point spectral indices similar to those calculated above would have to be flatter than our lower limits. Secondly, a long-lived neutron star merger remnant would be required. Whether such a stable remnant was, and remains, present, or collapsed to a black hole on a much shorter time-scale, has been the subject of considerable discussion in the literature \citep*[e.g.][]{metzger18,ai18,yu18,radice18b,piro19,gill19,yang19}.

\subsection{Forecasts for future LOFAR observations of GW events}\label{section:discussion2} 

The predicted afterglow light curve of GW170817 at 144 MHz is shown in the left panel of Figure~\ref{fig:modelling}. Here, we extrapolate the light curve at higher radio frequencies using a spectral index of $-$0.53 $\pm$ 0.04 \citep[][]{mooley18c}, corresponding to the assumption that both the characteristic synchrotron frequency, $\nu_{\rm m}$, and self-absorption frequency, $\nu_{\rm a}$, are lower than 144 MHz.\footnote{If this assumption fails, then the predicted flux density at 144 MHz is lower than that in Figure~\ref{fig:modelling}, and the peak of the light curve may be delayed; see e.g. \citet{nakar11}.} Indeed, following \citet[][]{hotokezaka16}, we find $\nu_{\rm a} \leq$ 36 MHz (for circum-merger density $n \leq$ 0.01 cm$^{-3}$, kinetic energy $E=$ 10$^{49}$ erg, fraction of internal energy given to the electrons $\epsilon_e=$ 0.1, fraction of internal energy contained in the magnetic field $\epsilon_B=$ 0.01, power-law index of the electron distribution $p=$ 2.2, and the initial velocity of the ejecta in units of the speed of light $\beta_{0}=$ 1; see discussion later in this section), which is well below the LOFAR HBA observing band. The aforementioned competing cocoon model, which has now been ruled out (see Section~\ref{section:introduction}) is also included. We use the cocoon model described in \citet[][]{mooley18a}: we use the kinetic energy distribution $E(>\gamma \beta) =$~2~$\times$~10$^{51}(\gamma \beta)^{-5}$ with $\gamma_{\rm max}=$ 3.5, where $\gamma$ and $\beta$ are the Lorentz factor and velocity, respectively. 

While our LOFAR upper limits lie well above the two curves, we can consider the hypothetical scenario of LOFAR late-time flux density measurements had an event similar to GW170817 been much further north on the sky. LOFAR can achieve a median noise level of approximately 70 $\upmu$Jy beam$^{-1}$ in routine 8-h observations, with 48 MHz bandwidth centred at 144 MHz, for declinations at or near zenith \citep[][]{shimwell19}. Considering the radio light curve fitting in \citet[][]{mooley18c} and also \citet[][]{alexander18} (in addition, see \citealt[][]{dobie18}), the peak flux density at 144 MHz would be predicted to be at approximately 7--9.5 times the median LOFAR sensitivity level (see left panel of Figure~\ref{fig:modelling}). Assuming that any uncertainties arising from a host galaxy contribution were negligible, this would have then allowed us to determine whether a single-power-law radio spectrum also held at low frequencies, or whether there were indications of spectral turnover. In the absence of spectral turnover, we would have also been able to discriminate at late times between competing models of the radio afterglow. 

GW170817 occurred relatively close by, and with a circum-merger density below average. It is beyond the scope of this paper to consider a comprehensive range of possible future compact binary mergers and their potential detectability with LOFAR. However, for illustrative purposes, let us now consider a more distant binary neutron star merger at 100 Mpc (i.e. about halfway to the Advanced LIGO design sensitivity horizon; \citealt[][]{abbott16}). As is shown in the right panel of Figure~\ref{fig:modelling}, the afterglows of jets, cocoons, and dynamical ejecta of such future merger events can be observed by LOFAR in certain cases if $n$ $\gtrsim$ 0.01 cm$^{-3}$. Note that observations of the afterglows of sGRBs show that 30--70 per cent of these events occur in environments where the density of the interstellar medium is $\gtrsim$ 0.01 cm$^{-3}$ \citep{fong15}. The relevant microphysical parameters are $\epsilon_{e}=$ 0.1, $\epsilon_{B}=$ 0.01, and $p=$ 2.2 (see \citealt[][]{hotokezaka16} for further details). We show a structured jet model, which has a uniform jet core up to a certain opening angle, and the energy and initial Lorentz factor decrease with angle as a power law. The kinetic energy of the jet core is 10$^{49}$ erg and the initial half-opening angle is 0.05 rad, with which the light curve is consistent with the observed features of the GW170817 afterglow \citep[][]{hotokezaka19}. We find that LOFAR can detect the afterglows of off-axis jets similar to GW170817 when the viewing angle is less than approximately 40\degr. The cumulative fraction of merger events detected by Advanced LIGO--Virgo with such a viewing angle is expected to be approximately one half \citep*{nissanke13}.  

In the right panel of Figure~\ref{fig:modelling}, we use a dynamical ejecta light curve (model `DNS$_{\rm m}$') taken from \citet[][]{hotokezaka16}. The dynamical ejecta may be partly responsible for the kilonova emission at optical and infrared wavelengths, but are unlikely to be the major component in terms of mass. However, since this component is faster than the disc outflow, i.e. the afterglow is brighter, we also consider the dynamical ejecta here.

In this discussion, we have assumed that the host galaxy flux density is negligible at LOFAR frequencies. This will not always be the case. Future GW events that are followed up by LOFAR will include an observation at roughly one week post merger, when early persistent emission and late-time afterglow emission are negligible. This observing strategy provides a comparison image to enable identification of the afterglow, but also a constraint on any host galaxy emission at the location of the GW event.

Future LOFAR observations will be particularly important to determine or constrain $\nu_{\rm a}$, which can be above 144 MHz (i.e. significantly higher than our calculation earlier in this section) and is sensitive to the velocity of the outflow \citep[e.g.][]{nakar11}. Measuring $\nu_{\rm a}$ will enable us to break the degeneracy between the model parameters, leading to a better estimate of the velocity and kinetic energy of the outflow. Not only will such measurements provide us with a better understanding of the afterglow, but will also help constrain the neutron star equation of state if the afterglow of the dynamical ejecta is detected \citep[e.g.][]{radice18a}.

\section{Conclusions and future work}\label{section:conclusions}

In this paper, we presented LOFAR follow-up observations of the compact binary merger event GW170817, which was detected by Advanced LIGO--Virgo. Our conclusions are as follows.  

\begin{enumerate}
\item In two sets of 4 $\times$ 2-h observations, occurring 130--138 and 371--374 days post merger, we determined $3\sigma$ upper limits of 6.6 (Run 1) and 19.5 (Run 2) mJy beam$^{-1}$ for the 144-MHz flux density of the electromagnetic counterpart, AT~2017gfo. 
\item Using previously published GMRT and ATCA flux densities at higher radio frequencies, we placed constraints on a number of two-point spectral indices for both AT~2017gfo, and the host galaxy NGC~4993, about 4.5 months post merger. In particular, for AT~2017gfo, $\alpha^{610}_{144} \gtrsim$ $-$2.5. The presence of ultra-steep-spectrum coherent radio emission at low frequencies would necessitate a long-lived neutron star remnant.
\item We showed that, for declinations at or near zenith, LOFAR will be able to detect various possible radio afterglows for a subset of future merger events.
\item We also demonstrated that it is possible to obtain images with LOFAR significantly south of the celestial equator, albeit a factor of about 1.5 dex less sensitive and at an angular resolution 2.5--5.3 times coarser than what is achievable at or near zenith, in this particular case. If LoTSS were to be extended below the celestial equator, with an angular resolution at or near the usual target value of 6 arcsec \citep[][]{shimwell17,shimwell19}, this would allow high-resolution, low-frequency sky models to be developed at declinations that will be readily accessible with the first phase of the low-frequency component of the Square Kilometre Array (i.e. SKA1--LOW). For example, \citet[][]{hale19} recently presented LOFAR HBA observations of the XMM Large-Scale Structure (XMM-LSS) field, which is centred at a declination of $-$4.5\degr. The angular resolution of their map is 8.5 arcsec $\times$ 7.5 arcsec, and the RMS noise level at the centre of the map is 280 $\upmu$Jy beam$^{-1}$. 
\end{enumerate}

Further LOFAR follow-up is planned for binary neutron star and black hole -- neutron star mergers that are detected in the current Advanced LIGO--Virgo observing run (`O3'). Follow-up will occur not only on the time-scales investigated in this paper, but also on time-scales as short as several minutes once an alert is received, using the LOFAR responsive telescope mode (see \citealt[][]{rowlinson19} for a review of the current rapid-response capabilities of a selection of low-frequency radio facilities, including the MWA and LWA). We can therefore expect to obtain further insight into the role that low-radio-frequency data will play in understanding the physical processes that occur following a compact binary merger containing at least one neutron star.   

\section*{Acknowledgements}

We thank the referee for a number of helpful suggestions that improved the content and presentation of this paper. We also thank the ASTRON Radio Observatory for setting up and scheduling the observations described in this paper, as well as preprocessing the data. Additionally, we are grateful to Ben Stappers for providing very useful feedback on an earlier version of this manuscript.

This paper is based on data obtained with the International LOFAR Telescope (ILT) under project codes DDT9\_002 and LT10\_013. LOFAR \citep[][]{vanhaarlem13} is the Low Frequency Array designed and constructed by ASTRON. It has observing, data processing, and data storage facilities in several countries, that are owned by various parties (each with their own funding sources), and that are collectively operated by the ILT foundation under a joint scientific policy. The ILT resources have benefitted from the following recent major funding sources: CNRS-INSU, Observatoire de Paris and Universit\'{e} d'Orl\'{e}ans, France; BMBF, MIWF-NRW, MPG, Germany; Science Foundation Ireland (SFI), Department of Business, Enterprise and Innovation (DBEI), Ireland; NWO, The Netherlands; The Science and Technology Facilities Council, UK; Ministry of Science and Higher Education, Poland.

The LOFAR direction-independent calibration pipeline (\url{https://github.com/lofar-astron/prefactor}) was deployed by the LOFAR e-infragroup on the Dutch National Grid infrastructure with support of the SURF Co-operative through grants e-infra170194, e-infra180087 and e-infra180169 \citep[][]{mechev17}. The LOFAR direction-dependent calibration and imaging pipeline (\url{http://github.com/mhardcastle/ddf-pipeline/}) was run on computing clusters at Leiden Observatory and the University of Hertfordshire, which are supported by a European Research Council Advanced Grant [NEWCLUSTERS-321271] and the UK Science and Technology Funding Council [ST/P000096/1].

PGJ acknowledges funding from the European Research Council under ERC Consolidator Grant agreement no. 647208. SC acknowledges funding support from the UnivEarthS Labex program of Sorbonne Paris Cit{\'e} (ANR-10-LABX-0023 and ANR-11-IDEX-0005-02). TMD acknowledges support from grants AYA2017-83216-P and RYC-2015-18148. JvL acknowledges funding from Vici research programme `ARGO' with project number 639.043.815, financed by the Netherlands Organisation for Scientific Research (NWO). MB acknowledges support by the Deutsche Forschungsgemeinschaft (DFG, German Research Foundation) under Germany's Excellence Strategy -- EXC 2121 ``Quantum Universe" -- 390833306.

This research has made use of the VizieR catalogue access tool, CDS, Strasbourg, France (DOI: 10.26093/cds/vizier). The original description of the VizieR service was published in A\&AS 143, 23 \citep*[][]{ochsenbein00}. This research has made use of the NASA/IPAC Extragalactic Database (NED), which is operated by the Jet Propulsion Laboratory, California Institute of Technology, under contract with the National Aeronautics and Space Administration. This research has made use of NASA's Astrophysics Data System Bibliographic Services. This project also made use of {\sc kvis} \citep[][]{gooch95}, {\sc topcat} \citep[][]{taylor05}, {\sc numpy} \citep[][]{oliphant06} and {\sc matplotlib} \citep[][]{hunter07}.





\noindent
{\textit{
\newline
$^{1}$ASTRON, the Netherlands Institute for Radio Astronomy, Oude Hoogeveensedijk 4, 7991 PD Dwingeloo, The Netherlands\\
$^{2}$International Centre for Radio Astronomy Research, Curtin University, GPO Box U1987, Perth, WA 6845, Australia\\
$^{3}$Leiden Observatory, Leiden University, PO Box 9513, 2300 RA, Leiden, The Netherlands\\
$^{4}$Anton Pannekoek Institute for Astronomy, University of Amsterdam, Science Park 904, 1098 XH Amsterdam, The Netherlands\\
$^{5}$GRAPPA, Anton Pannekoek Institute for Astronomy and Institute of High-Energy Physics, University of Amsterdam, Science Park 904, 1098 XH Amsterdam, The Netherlands\\
$^{6}$Nikhef, Science Park 105, 1098 XG Amsterdam, The Netherlands\\
$^{7}$Department of Astrophysical Sciences, Peyton Hall, Princeton University, Princeton, NJ 08544, USA\\
$^{8}$Research Center for the Early Universe, Graduate School of Science, University of Tokyo, Bunkyo-ku, Tokyo 113-0033, Japan\\
$^{9}$SRON Netherlands Institute for Space Research, Sorbonnelaan 2, 3584 CA, Utrecht, The Netherlands\\
$^{10}$Department of Astrophysics/IMAPP, Radboud University Nijmegen, PO Box 9010, 6500 GL Nijmegen, The Netherlands\\
$^{11}$GEPI and USN, Observatoire de Paris, Universit{\'e} PSL, CNRS, 5 Place Jules Janssen, 92190 Meudon, France\\
$^{12}$Department of Physics and Electronics, Rhodes University, PO Box 94, Grahamstown, 6140, South Africa\\
$^{13}$Centre for Astrophysics Research, School of Physics, Astronomy and Mathematics, University of Hertfordshire, College Lane, Hatfield AL10 9AB, UK\\
$^{14}$SURFsara, PO Box 94613, 1090 GP Amsterdam, The Netherlands\\
$^{15}$Astrophysics, Department of Physics, University of Oxford, Keble Road, Oxford OX1 3RH, UK\\
$^{16}$Sydney Institute for Astronomy, School of Physics, The University of Sydney, NSW 2006, Australia\\
$^{17}$CSIRO Astronomy and Space Science, PO Box 76, Epping NSW 1710, Australia\\
$^{18}$Department of Physics, University of Virginia, PO Box 400714 Charlottesville, Virginia 22904-4714, USA\\
$^{19}$Institute of Astronomy, University of Cambridge, Madingley Road, Cambridge CB3 0HA, UK\\
$^{20}$University of Technology Sydney, 15 Broadway, Ultimo NSW 2007, Australia\\
$^{21}$University of the Virgin Islands, 2 Brewers Bay Road, Charlotte Amalie, USVI 00802, USA\\
$^{22}$Laboratoire AIM (CEA/IRFU - CNRS/INSU - Universit{\'e} Paris Diderot), F-91191 Gif-sur-Yvette, France\\
$^{23}$Station de Radioastronomie de Nan\c{c}ay, Observatoire de Paris, PSL Research University, CNRS, Univ. Orl\'{e}ans, OSUC, 18330 Nan\c{c}ay, France\\
$^{24}$Th{\"u}ringer Landessternwarte, Sternwarte 5, D-07778 Tautenburg, Germany\\
$^{25}$LPC2E - Universit\'{e} d'Orl\'{e}ans /  CNRS, 45071 Orl\'{e}ans cedex 2, France\\
$^{26}$SKA Organisation, Jodrell Bank Observatory, SK11 9DL, UK\\
$^{27}$Cahill Center for Astronomy and Astrophysics, California Institute of Technology, Pasadena, CA, USA\\
$^{28}$Instituto de Astrof{\'i}sica de Canarias, 38205 La Laguna, Tenerife, Spain\\
$^{29}$Departamento de Astrof{\'i}sica, Universidad de La Laguna, E-38206 La Laguna, Tenerife, Spain\\
$^{30}$South African Radio Astronomy Observatory, 2 Fir Street, Black River Park, Observatory, 7925, South Africa\\
$^{31}$Department of Physics and Astronomy, University of the Western Cape, Cape Town 7535, South Africa\\
$^{32}$Department of Physics, The George Washington University, 725 21st Street NW, Washington, DC 20052, USA\\
$^{33}$Astronomy, Physics, and Statistics Institute of Sciences (APSIS), The George Washington University, Washington, DC 20052, USA\\
$^{34}$LESIA and USN, Observatoire de Paris, CNRS, PSL, SU/UP/UO, 92195 Meudon, France\\
$^{35}$Technical University of Berlin, Institute of Geodesy and Geoinformation Science, Faculty VI sec. H 12, Main Building Room H 5121, Stra{\ss}e des 17. Juni 135, 10623 Berlin, Germany\\
$^{36}$GFZ German Research Centre for Geosciences, Telegrafenberg, 14473 Potsdam, Germany\\
$^{37}$Eindhoven University of Technology, PO Box 513, 5600 MB Eindhoven, The Netherlands\\
$^{38}$Kapteyn Astronomical Institute, University of Groningen, PO Box 800, 9700 AV Groningen, The Netherlands\\
$^{39}$Hamburger Sternwarte, University of Hamburg, Gojenbergsweg 112, D-21029 Hamburg, Germany\\
$^{40}$Max Planck Institute for Astrophysics, Karl-Schwarzschild-Str. 1, 85748 Garching, Germany\\
$^{41}$Jodrell Bank Centre for Astrophysics, Alan Turing Building, Department of Physics and Astronomy, The University of Manchester, Oxford Road, Manchester M13 9PL, UK\\
$^{42}$Vrije Universiteit Brussel, Department of Physics and Astronomy, B-1050 Brussels, Belgium\\
$^{43}$Astronomisches Institut der Ruhr-Universit{\"a}t Bochum, Universit{\"a}tsstra{\ss}e 150, 44780 Bochum, Germany\\
$^{44}$Space Radio-Diagnostics Research Centre, University of Warmia and Mazury in Olsztyn, Prawoche{\'n}skiego 9, 10-720 Olsztyn, Poland\\
$^{45}$Leibniz-Institut f{\"u}r Astrophysik Potsdam (AIP), An der Sternwarte 16, D-14482 Potsdam, Germany\\
$^{46}$Deutsches Elektronen-Synchrotron (DESY), 15738 Zeuthen, Germany\\
$^{47}$Erlangen Centre for Astroparticle Physics (ECAP), Friedrich-Alexander-Universit{\"a}t Erlangen-N{\"u}rnberg, 91058 Erlangen, Germany\\
$^{48}$Center for Information Technology (CIT), University of Groningen, Groningen, The Netherlands\\
$^{49}$Univ. Lyon, Univ. Lyon1, Ens de Lyon, CNRS, Centre de Recherche Astrophysique de Lyon UMR5574, 9 av. Charles Andr{\'e} F-69230, Saint-Genis-Laval, France\\
$^{50}$Pozna{\'n} Supercomputing and Networking Center (PCSS), Pozna{\'n}, Poland\\
$^{51}$Max-Planck-Institut f{\"u}r Radioastronomie, Auf dem H{\"u}gel 69, 53121 Bonn, Germany\\
$^{52}$Fakult{\"a}t f{\"u}r Physik, Universit{\"a}t Bielefeld, Postfach 100131, 33501 Bielefeld, Germany\\
$^{53}$Jagiellonian University, Astronomical Observatory, ul. Orla 171, 30-244 Krak{\'o}w, Poland\\
$^{54}$Department of Space, Earth and Environment, Chalmers University of Technology, Onsala Space Observatory, SE-439 92 Onsala, Sweden\\
}}


\bsp	
\label{lastpage}
\end{document}